# Impact of Nitrogen Atom Clusters and Vacancy Defects on Graphene: A Molecular Dynamics Investigation

Indranil Rudra[1], Md. Moktadir Billah Tahmid[1], Jahid Emon[2*], Mohammad Jane Alam Khan[1]

[1]Department of Mechanical Engineering, Bangladesh University of Engineering and Technology, Dhaka, Bangladesh

[2]Department of Mechanical Science & Engineering, University of Illinois Urbana-Champaign, Urbana, IL, United States

[*]Corresponding authors: jemon2@illinois.edu

## Abstract

Graphene's exceptional mechanical properties are crucial for its integration into advanced technological applications. However, real-world synthesis and functionalization processes introduce structural modifications that can compromise its mechanical integrity. Nitrogen doping, while beneficial for electronic property tuning, often results in atomic clustering rather than uniform distribution, while concurrent vacancy defect formation represents another common structural alteration during processing. This study systematically investigates the comparative effects of nitrogen atom clusters and equivalent-sized vacancy defects on the mechanical behavior of graphene sheets through molecular dynamics simulations. The Nitrogen clustering significantly degraded mechanical performance almost similarly to random doping. In comparison, systems with equivalent-sized vacancy defects showed higher stiffness and lower ductility than those with clusters. The study revealed distinct failure mechanisms between doped and defective configurations, with nitrogen clusters showing modified crack propagation patterns while vacancies acted as pronounced stress concentrators, leading to premature failure. However, this study also showed that defect morphology critically influences mechanical properties. These findings provide important insights for optimizing graphene synthesis and processing protocols, highlighting the differential mechanical risks associated with dopant clustering versus vacancy formation. The results inform defect-tolerant design strategies for graphene-based nanoelectronics, composites, and sensors, where mechanical reliability is paramount for device performance and longevity.

# Introduction

Graphene, a single-layer arrangement of sp²-hybridized carbon atoms in a hexagonal lattice, exhibits remarkable mechanical properties including ultra-high strength [1], exceptional stiffness [1, 2], and extraordinary flexibility [3]. These superior mechanical characteristics, combined with excellent electrical and thermal properties [4, 5], position graphene as a cornerstone material for next-generation technologies spanning flexible electronics [6], high-performance composites [7], energy storage devices [8], and advanced sensors [9]. However, the translation of graphene from laboratory demonstrations to practical applications necessitates controlled modification of its intrinsic properties through doping and functionalization processes [6, 10]. Nitrogen doping represents one of the most extensively studied approaches for tailoring graphene's electronic properties while preserving much of its structural integrity [11-13]. The incorporation of nitrogen atoms into the carbon lattice can be achieved through various synthesis methods including chemical vapor deposition (CVD) with nitrogen-containing precursors [14], post-synthesis plasma treatment [15], and thermal annealing in nitrogen-rich environments [16]. However, the atomic-scale distribution of nitrogen dopants significantly influences both electronic and mechanical properties, with clustering effects often observed under realistic synthesis conditions [17].

The formation of nitrogen clusters, rather than uniform substitutional doping, arises from thermodynamic considerations and kinetic limitations during synthesis [18]. While nitrogen clustering can create unique electronic states and enhance specific functionalities such as electrocatalytic activity [19], the mechanical implications of such clustering remain inadequately understood. Concurrent with intentional doping processes, vacancy defects frequently form due to carbon atom removal during synthesis [20], plasma exposure [21], or chemical functionalization [22], creating holes that may be comparable in size to dopant clusters. Molecular dynamics (MD) simulations have emerged as powerful tools for investigating the mechanical behavior of graphene at the atomistic level [22], providing insights into deformation mechanisms [23], failure modes, and structure-property relationships [24] that are difficult to probe experimentally. The Tersoff potential, widely validated for carbon-based systems [25], offers a robust framework for describing the bonding environment in both pristine [26] and modified graphene structures [27]. Previous MD studies have primarily focused on the effects of isolated point defects [28], random doping [27], or simple vacancy arrays [22], leaving a significant knowledge gap regarding the comparative mechanical impact of clustered defects versus extended vacancy regions.

The mechanical integrity of graphene-based materials is often the limiting factor in device reliability and performance, particularly in applications involving mechanical stress or strain. Understanding how different types of structural modifications, specifically nitrogen clusters versus vacancy defects of equivalent size, affect key mechanical properties such as Young's modulus, Ultimate tensile strength, toughness, and fracture strain is crucial for informed material design. This knowledge is particularly relevant given that both defect types can occur simultaneously during synthesis and processing, yet their relative mechanical impacts and underlying failure mechanisms may differ substantially. Recent experimental studies have highlighted the critical role of defect size [29] and distribution [28, 30] in determining the mechanical response of two-dimensional materials. Theoretical investigations have suggested that the mechanical degradation in defective graphene depends not only on defect concentration but also on defect morphology [31, 32], local stress concentration factors [26, 30], and the specific atomic arrangements around defect sites [20].

However, systematic comparative studies examining equivalent-sized nitrogen clusters versus vacancy defects under identical loading conditions remain limited in the literature. The present study addresses this knowledge gap by conducting comprehensive molecular dynamics simulations to systematically compare the mechanical effects of nitrogen atom clusters and equivalent vacancy defects in graphene sheets. By employing a rigorous computational framework with validated interatomic potentials and standardized mechanical testing protocols, this work aims to elucidate the fundamental relationships between defect type, size, and mechanical performance. The insights gained from this investigation will inform the development of defect-tolerant graphene materials and guide synthesis optimization for applications where mechanical reliability is paramount.

## Methodology

### *Atomistic Model and Simulation Setup*

All molecular dynamics (MD) simulations were performed using the Large-scale Atomic/Molecular Massively Parallel Simulator (LAMMPS) package [33], with initial atomic configurations generated by using atomsk [34]. A pristine, defect-free single layer graphene sheet was first constructed with dimensions of approximately 49.2 Å × 85.2 Å, corresponding to the zigzag (x) and armchair (y) directions, respectively, containing a total of 1600 carbon atoms. Periodic boundary conditions were applied along the planar x and y directions, with a fixed boundary in the z-direction. Two types of circular defects, with radius ranging from 4 Å to 10 Å, were systematically introduced at the geometric center of the sheet. One type of defect was substitutional nitrogen (N-doping), where carbon atoms within the specified radius were replaced by nitrogen. The other type was vacancy defect, where carbon atoms within the radius were removed. The interactions between carbon (C) and carbon (C), and carbon (C) nitrogen (N) atoms were modeled using the bond-order Tersoff potential [25]. This potential was chosen for its ability to accurately describe the covalent bonding and bond-breaking essential for simulation of material fracture. In the Tersoff potential, the total energy of the system is given by:

$$E = \frac{1}{2} \sum_i \sum_{j \neq i} V_{ij} \tag{1}$$

$$V_{ij} = f_C(r_{ij})[f_R(r_{ij}) + b_{ij} f_A(r_{ij})] \tag{2}$$

Where $r_{ij}$ is the distance between the pairs of adjacent atoms i and j, and $b_{ij}$ is an empirical bond-order coefficient, $f_R$ represents the repulsive pair-wise interaction, while $f_A$ stands for the attractive term, which are given as follows:

$$f_R(r) = A \exp(-\lambda_1 r) \tag{3}$$
$$f_A(r) = -B \exp(-\lambda_2 r) \tag{4}$$

Where A, B, $\lambda_1$, and $\lambda_2$ represent repulsive pair potential prefactor, Attractive pair potential prefactor, decay length for repulsive term, and decay length for attractive term respectively. The cutoff function $f_C(r)$ ensures smooth decay of interactions which is expressed as:

$$f_C(r) = \begin{cases} 1 & r < R - D \\ \frac{1}{2} - \frac{1}{2} \sin\left(\frac{\pi}{2} \frac{r-R}{D}\right) & R - D < r < R + D \\ 0 & r > R + D \end{cases} \tag{5}$$

Where R is the cutoff distance, and D is the Cutoff width.

The bond order term $b_{ij}$ captures the short-range local atomic environment which is expressed as:

$$b_{ij} = (1 + \beta^n \zeta_{ij}{}^n)^{-\frac{1}{2n}} \tag{6}$$

$$\zeta_{ij} = \sum_{k \neq i,j} f_C(r_{ik}) g(\theta_{ijk}) \exp[\lambda_3{}^m (r_{ij} - r_{ik})^m] \tag{7}$$

Where $\lambda_3, \beta, \text{n, and m}$ are the decay length for angular term, angular strength parameter, angular exponent, and bond order exponent respectively. $\theta_{ijk}$ is the angle between bonds $ij$ and $ik$. The angular function $g(\theta)$ is:

$$g(\theta) = \gamma \left(1 + \frac{c^2}{d^2} - \frac{c^2}{d^2 + (h - \cos\theta)^2}\right) \quad (8)$$

Where c, d, and $\gamma$ are the angular coefficient, angular parameter, and angular scaling respectively. Moreover, $h = \cos\theta_0$ defines the preferred angle $\theta_0$.

*Simulation*

A multi-step procedure was employed to equilibrate the structures and perform the tensile tests. Firstly, an energy minimization was performed using the conjugate gradient (CG) algorithm to remove any unfavorable atomic overlaps. For our primary simulations, the system was relaxed until the energy tolerance reached $1\times10^{-11}$ eV or the force tolerance reached $1\times10^{-6}$ eV/Å. The robustness of this system was confirmed through a parallel set of validation simulations using stricter minimization criteria (energy tolerance of $1\times10^{-12}$ eV and force tolerance of $1\times10^{-8}$ eV/Å), which yielded negligible differences in the final mechanical properties. Following minimization, the system was thermally equilibrated within the isothermal-isobaric (NPT) ensemble for 5 ps at 300 K and zero pressure, using a Nosé-Hoover thermostat [35, 36] and barostat with a fine timestep of 0.1 fs. After equilibration, uniaxial tensile loading was simulated by deforming the simulation box along the y-direction (armchair) at a constant engineering strain rate of $1\times10^9$ s$^{-1}$. During deformation, the temperature was maintained at 300 K and the pressure in the transverse x-direction was kept at zero to allow for natural contraction due to the Poisson effect. The tensile test was run for a total of 0.25 ns, corresponding to a strain of at least 25%, to ensure full capture of the fracture event, with atomic data dumped every 0.5 ps for post-processing.

*Analysis of Mechanical Properties*

The mechanical properties were derived from the simulation outputs. The stress tensor for the system was calculated based on the Virial theorem, and the 3D engineering stress was determined by normalizing the relevant tensor component with the instantaneous volume of the simulation box, assuming an effective graphene thickness of 3.4 Å. Stress-strain curves were then plotted for each structure. From these curves, key properties including Young's Modulus, Ultimate tensile strength, and fracture strain were extracted. Visualization and further post-processing of the atomic trajectories were performed using the OVITO software [37].

## Results and Discussion

*Validation of MD simulations*

To ensure the accuracy and reliability of the molecular dynamics (MD) simulation methodology, a foundational tensile test was conducted on a pristine graphene system. Interatomic interactions were accurately described using the Tersoff potential, a robust choice for systems containing carbon and nitrogen atoms.

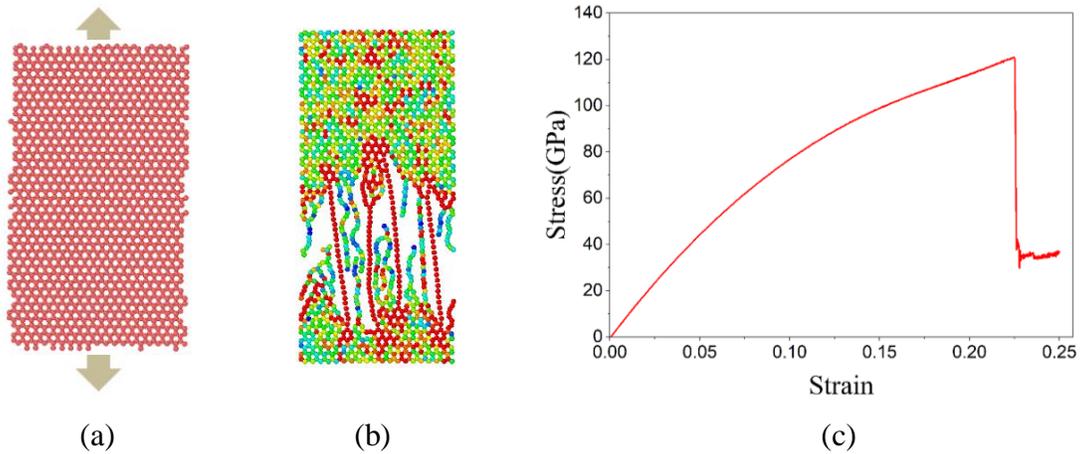

(a)          (b)          (c)

**Figure 1.** (a) MD configuration, (b) post-fracture configuration (c) stress-strain curve of pristine graphene.

The simulated system comprised 1600 carbon atoms, subjected to uniaxial tensile loading (Figure 1(a)). A stress-strain curve was generated shown in Figure 1(b), revealed a Young's modulus of approximately 969.97 GPa, an Ultimate tensile strength of 120.58 GPa, and a fracture strain of 0.226.

**Table 1:** Comparison of mechanical properties between MD simulations of present study and existing literature.

| Property | Current Study | Mortazavi et al. [27] | Ansari et al. [22] |
| --- | --- | --- | --- |
| Young's Modulus (GPa) | 969.97 | 996 | -- |
| Ultimate tensile strength (GPa) | 120.58 | -- | 123 |
| Fracture Strain | 0.226 | 0.22 | 0.233 |

As shown in Table 1, all properties (Young's modulus, ultimate tensile strength, fracture strain) of our study agree well with values reported by Mortazavi et al. and Ansari et al., confirming that the simulation protocol (1600-atom graphene, 300 K, Tersoff potential) reliably reproduces established mechanical behaviors of pristine graphene. This strong validation ensures confidence in subsequent defect-impact analyses.

*Effects of Nitrogen Cluster Doping on Graphene's Mechanical Properties*

Incorporation of nitrogen clusters into the graphene lattice induces a systematic degradation of mechanical performance. Using the data from stress-strain curve (Figure 2(a-b)) the trends (Figure 2(c-e)) are plotted. With the increasing size of cluster, Young's modulus decreases nearly linearly from ~966 GPa (4 Å) to ~805 GPa (10 Å) which is shown in Figure 2(d). This reflects local bond weakening and lattice distortion around N-rich regions [38], consistent with reactive-MD results showing that N dopants disrupt sp² C–C bonding [20] and introduce ripples [39] that lower stiffness. Figure 2(e) shows that ultimate tensile strength drops from 112.09 GPa to 70.36 GPa. Fracture strain falls from 0.1961 to 0.1217 (Figure 2(e)), indicating reduced ductility. The behavior parallels trends in substitutional N-doping studies reported by H. Ghorbanfekr-Kalashami et al. [28], where fracture strain decreases nearly exponentially with dopant concentration. The single N atom present in the C ring is referred to pyridinic-N, and pyrazole-N is characterized by having two adjacent nitrogen atoms in the C ring. In the smallest cluster all the N atoms are pyridinic tightly bonded with C atoms, exhibiting much higher mechanical properties than the graphene sheets having larger cluster. The number of C-N bonds (and N-N bonds) increases and there is an increased chance to have two C-N bonds adjacent to each other. The latter results in a nonlinear increasing disturbance in the C-N, C-C, and N-N bond lengths [28]. In smaller sized clusters, pyridinic-N configuration is obtained (Figure 3(a)). Pyridinic N at an edge or vacancy actively alters both local bond angles and charge density [20]. It can blunt stress concentration, delay crack initiation, and allow slightly greater elastic stretch before fracture [38]. The latter systems consist of pyrazole-N (Figure 3(b)) instead of pyridinic-N configuration. It is noticed from the simulations that the hexagons of C having pyrazole-N acts as crack initiation site. A modest uptick in case of 8 Å cluster suggests size-dependent crack-defect interaction. Previous MD investigations have shown that defect morphology critically influences fracture strain. In this system, the arrangement of N atoms bonded to C atoms at the defect's edge is in symmetric, isotropic fashion which may result in higher ultimate tensile strength, and fracture strain compared to more anisotropic clustered defect distributions that is present in the other systems.

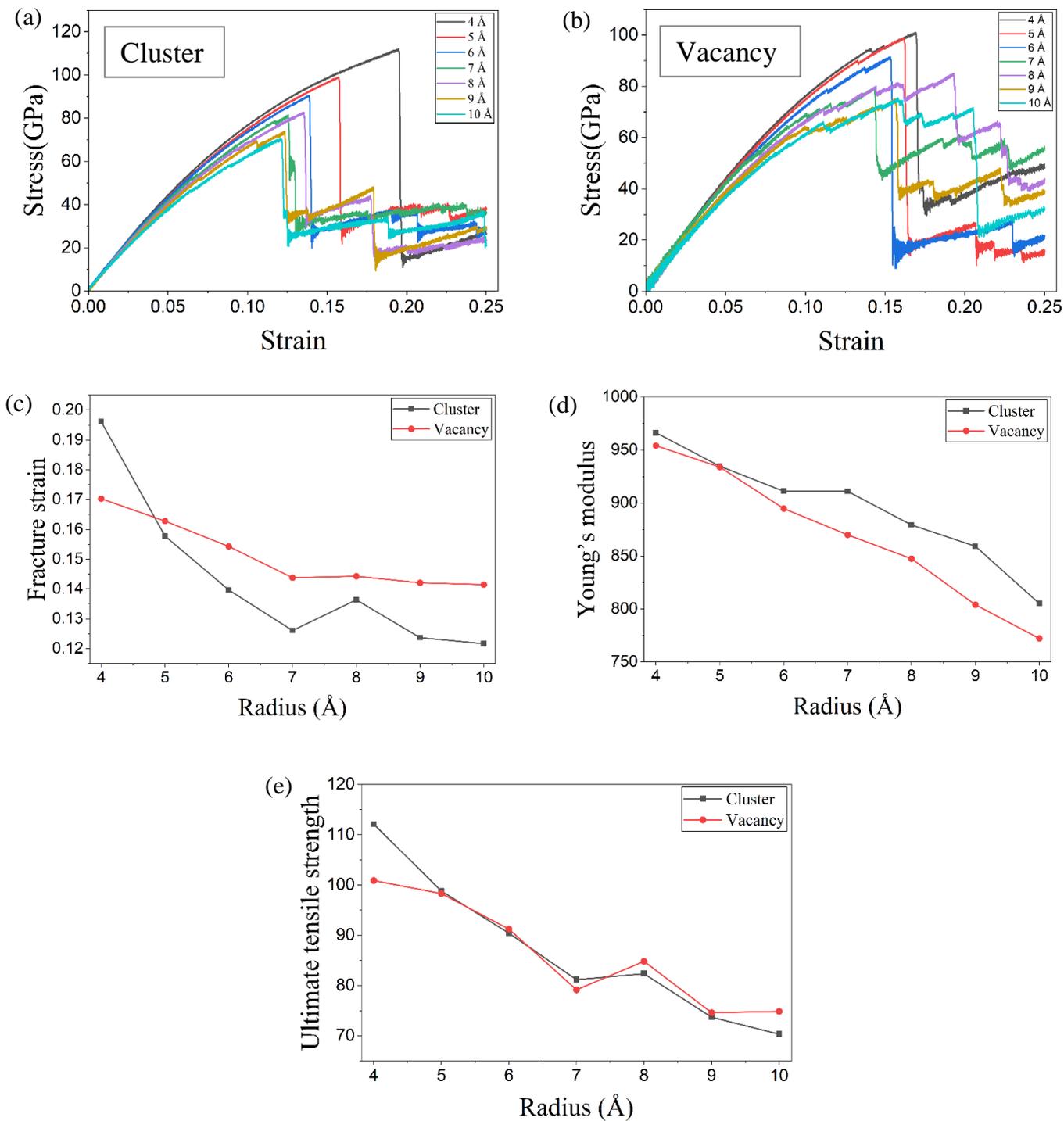

**Figure 2.** Stress-strain curves for (a) clusters (b) vacancy, and variation of (c) fracture strain (d) Young's modulus (e) ultimate tensile strength of graphene sheet having clusters, and vacancy defects with radius.

*Effects of circular vacancy defects on Graphene's Mechanical properties*

Young's modulus demonstrates a nearly linear decline as the size of the circular vacancy increases. For the graphene sheet having the smallest vacancy this value approaches ~954 GPa; but with the largest vacancy (10 Å hole), the modulus drops to approximately ~772 GPa (Figure 2(d)) This decrease arises from the removal of strong sp² C–C bonds [20] and the loss of lattice continuity, which collectively diminish the in-plane stiffness of the sheet [39]. Larger holes disrupt more hexagonal rings and aggravate lattice softening, consistent with simulation and experimental reports [27]. ultimate tensile strength drops sharply as vacancy size increases. The value drops from 100.89 GPa for the smallest hole to a value of 74.90 GPa for the largest hole (Figure 2(e)). Circular vacancies serve as major stress concentrators [20]. Under tensile loading, local stress accumulates around vacancy edges, facilitating early bond rupture and premature failure as described in molecular dynamics studies [27, 39]. Fracture strain declines as vacancies grow, from ~0.17 down to ~0.14 and stabilizing at the lowest values for large defects (Figure 2(c)). A slight increase in the value of fracture strain, and ultimate tensile strength for the vacancy of radius 8 Å is noticed (Figure 2(c)). The vacancy here is nearly symmetric, and relatively smooth-edged (Figure 3(d)). As a result, crack initiation does not occur at the edge of the main hole but emerges in adjacent lattice regions. This is a distinct contrast to systems with smaller or less symmetric vacancies (Figure 3(c)), where cracks typically nucleate and propagate directly from high-stress locations at the vacancy rim. The absence of sharp notches or jagged features distributes local stresses more evenly around the rim of the hole. Consequently, the stress concentration near the vacancy is reduced compared to highly irregular defects, resulting in higher ultimate tensile strength, and fracture strain than the other configurations.

## *The comparative effects of N atom clusters and equivalent-sized vacancy defects on Graphene's Mechanical Properties*

For the graphene sheet having 4 Å cluster, 3 out of 4 atoms are pyridinic-N, which apparently gives higher values of mechanical properties than that having 4 Å hole. If we look up to the rest of the systems, graphene with vacancy defects is noticeably more ductile than with N atom clusters. For vacancy defects the rounded carbon rim can stretch slightly further before break, whereas hetero-atom rich edges in clusters may act as earlier bond-rupture sites once the chemical crack-blunting benefit is exhausted. However, clusters conserve some load-bearing connectivity via N–C bonds, so Young's modulus is seen to exhibit higher value than equivalent-sized vacancy defects. So, if stiffness retention is the priority, nitrogen cluster doping is preferable across most sizes. To find out the fundamental reason behind the superior stiffness retention in N-doped graphene compared to sheets with vacancies, we conducted a quantitative analysis of the local atomic structure and energetics at the defect sites. The bond lengths and per-atom potential energies for the equilibrated structures were calculated. The average C-N (1.414 Å) and adjacent C-C (1.425 Å) bond lengths, and their low standard deviations (which are 0.051 and 0.035 respectively) indicate a uniform, low-strain structure. This structural integrity is powerfully reflected in the local energetics; the average per-atom potential energy in the atoms near the N-cluster is -7.35 eV, which confirms a very stable, low-energy configuration. In contrast, the vacancy introduces a profound local structural and energetic disruption. The C-C bonds at the vacancy edge are highly distorted, evidenced by a massive standard deviation of 0.236 Å. This severe structural instability results in a significant energetic penalty, with the edge atoms having an average potential energy of -6.82 eV. This value confirms that these atoms exist in a high-energy, unstable state due to unsaturated bonds and extreme local strain. In case of the value of ultimate tensile strength, clustered system shows higher value than vacancy defected system for smaller sizes. With further increase in the size, configurations with vacancy defects exhibit higher value.

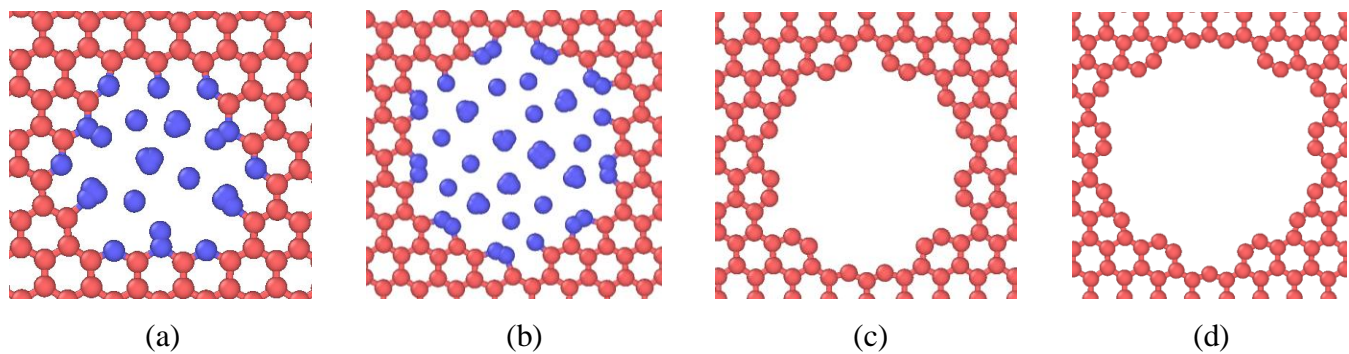

(a)          (b)          (c)          (d)

**Figure 3.** Representative atomic structures of graphene sheets illustrating the presence of cluster of (a) 6 Å (b) 8 Å radius, and hole of (c) 7 Å (d) 8 Å radius

*Fracture Mechanism of Graphene with Clustered and Vacancy Defects*

The system with 4 Å cluster didn't show any particular region of stress concentration as most of the N atoms were in bond with C atoms. Ligament mesh was seen in the middle of the sheet and were parallel to the loading direction (Figure 4(a),4(h)). With further increase in load, the ligament from sides began to rupture first and slowly proceeded toward the middle too. In the medium sized clustered (5 Å- 6 Å) sheets, the fracture initiates by breaking the hexagons of C atoms only (Figure 4(b-c), 4(i-j)). No significant quantity of ligaments is seen here. However, because of the presence of pyridinic N in 6 Å clustered sheet which blunts stress concentration the crack initiation is delayed. For the larger clustered (7 Å- 10 Å) sheets the crack is initiated in the pyrazole-N sites (Figure 4(d-g), 4(k-n)). Short chained ligaments are dominant here. Crack initiation is also delayed for 8 Å clustered system as the arrangement of N atoms at the edge is in symmetric, isotropic fashion. The failure mode remains fundamentally brittle and catastrophic across all sizes. While evidence of minor energy dissipation mechanisms like crack-tip blunting emerges, these are generally insufficient to overcome the negative effects of increased stress concentration from irregular geometries, often resulting in lower fracture strain for larger, asymmetric clusters. In cases of the smaller holed graphene sheets (4 Å- 5 Å) the hole is bridged by a few well-defined atomic ligaments (Figure 5(a-b), 5(h-j)). Fracture is clean and direct. The post-fracture images show that the sheet has snapped along a relatively straight path running through the vacancy. The failure is abrupt and catastrophic, characteristic of classic brittle materials. Intermediate vacancy defect (6 Å) shows a different scenario due to its structure (Figure 5(e),5(j)). The perimeter of the hole has a uniform shape with no noticeable concavities. This leads to more uniform ligament formation and breaking. As the vacancy size increases, the fracture path becomes more intricate (Figure 5(d-g), 5(k-n)). For these large vacancies, the hole is spanned by a web-like network of atomic strands rather than a few simple ligaments. When a tensile force is applied to the sheet it gets funneled along these preferential directions, creating these stress concentration bands that emanate from the defect's edge. These bands represent the paths of least resistance for energy transfer through the lattice. However, with a larger hole the stress concentration zone is more spread out. This can lead to the formation of multiple, less intensely defined stress bands around the hole. Instead of one overwhelmingly dominant diagonal path, there might be several potential pathways for crack propagation and the stress might distribute more broadly. However, the failure in the systems containing medium sized defects such as 7 Å vacancy propagates by unzipping along one pre-weakened dominant diagonal path, which results in a clean, straight fracture that follows the stress band (Figure 5(d),5(k)). For same reason, similar post fracture scenario is noticed for same sized (7 Å rad) clustered system (Figure 4(k)).

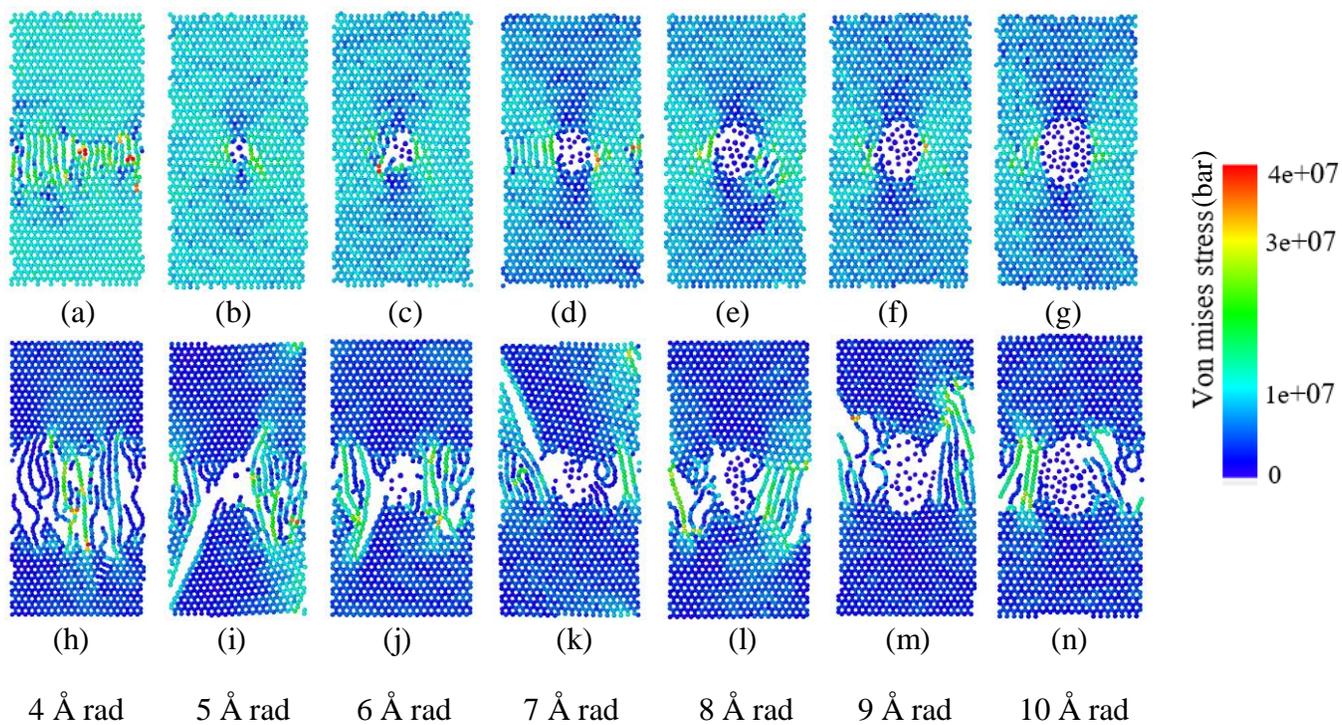

**Figure 4.** Atomic structures of graphene sheets (a-g) before, and (h-n) after the fracture with clusters.

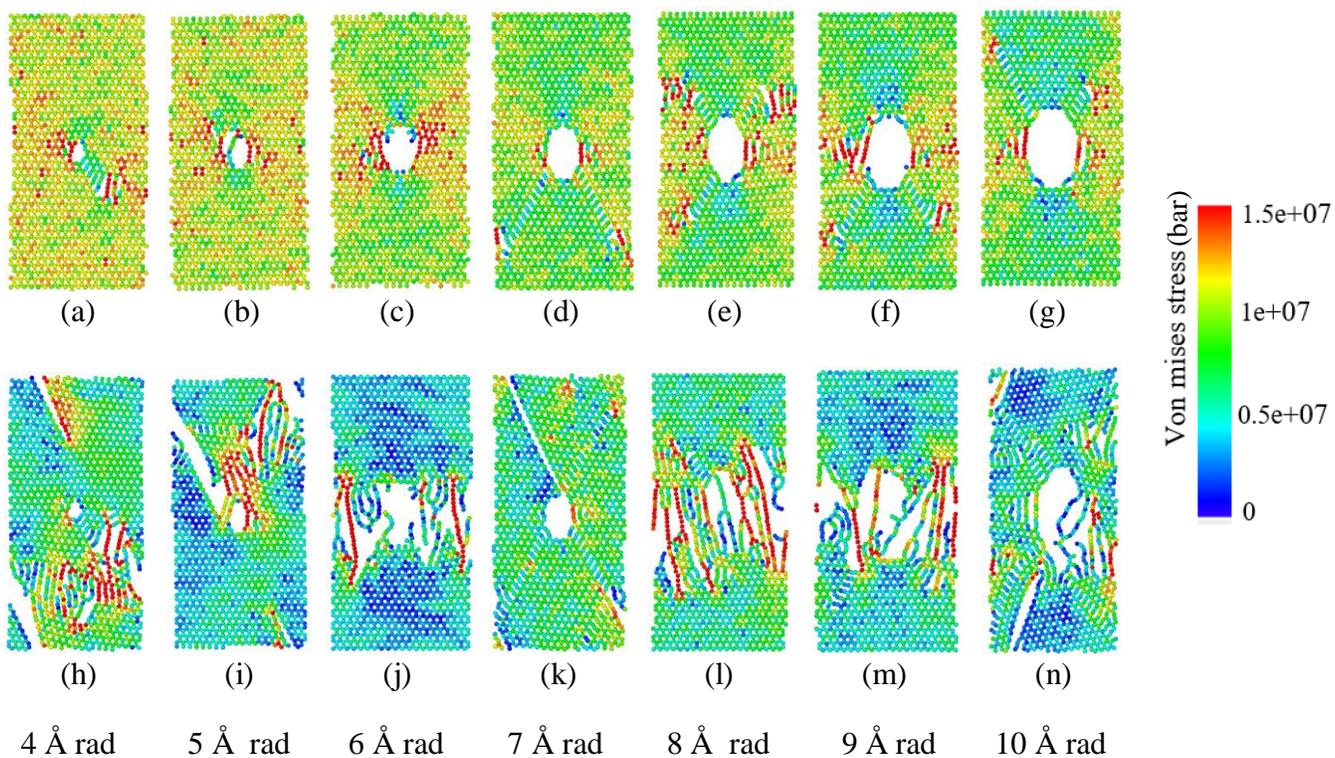

**Figure 5.** Atomic structures of graphene sheets (a-g) before, and (h-n) after the fracture with vacancy defects.

## Conclusion

This study used molecular dynamics simulations to investigate how nitrogen atom clusters and equivalent-sized vacancy defects affect graphene's mechanical behavior. Both nitrogen clustering and vacancy formation significantly degraded graphene's mechanical performance compared to pristine graphene. Nitrogen clusters showed modified crack propagation patterns. Smaller clusters led to ligament formation and rupture, while medium-sized clusters showed fracture initiating by breaking carbon hexagons. Larger clusters exhibited crack initiation at pyrazole-N sites, involving short-chained ligaments. Smaller vacancies resulted in clean, direct fractures through bridging ligaments. Larger vacancies led to web-like networks of atomic strands and more distributed stress concentration. The symmetry of defect edges significantly influenced crack initiation and propagation, sometimes leading to unexpected increases in properties for specific defect sizes. Vacancy defected graphene was generally more ductile than nitrogen-clustered graphene in our study. Whereas, clustered graphene showed higher stiffness than configuration with equivalent sized vacancy defects. These findings offer important insights for optimizing graphene synthesis by highlighting the differential mechanical risks associated with dopant clustering versus vacancy formation, which are both common during graphene modification. The comparative study can be extended to other common dopants (e.g., boron) and functionalization groups to build a more comprehensive understanding of how different chemical modifications affect mechanical integrity. Moreover, correlation of mechanical properties with electrical, thermal, and chemical reactivity traits can be done to guide multifunctional device optimization.

## Data availability

Data will be made available on request.